\newcommand*{\ATLASLATEXPATH}{}
\documentclass[cernpreprint,texlive=2011,txfonts,UKenglish]{atlasdoc}
\pdfoutput=1
\usepackage[biblatex=false]{\ATLASLATEXPATH atlaspackage}
\usepackage{\ATLASLATEXPATH atlascontribute}

\usepackage{\ATLASLATEXPATH atlasphysics}
\graphicspath{{}{figures/}}

\hypersetup{pdftitle={ATLAS draft},pdfauthor={The ATLAS Collaboration}}
\AtlasTitle{Measurement of the branching ratio
$\Gamma(\Lambda_b^0 \rightarrow \psi(2S)\Lambda^0)/\Gamma(\Lambda_b^0 \rightarrow J/\psi\Lambda^0)$
with the ATLAS detector}

\author{The ATLAS Collaboration}

\date{\today}

\AtlasRefCode{BPHY-2013-08}

\PreprintIdNumber{CERN-PH-EP-2015-164}

\AtlasJournal{Phys.\ Lett.\ B.}
\AtlasAbstract{%
An observation of the $\Lambda_b^0 \rightarrow \psi(2S) \Lambda^0$ decay
and a comparison of its branching fraction
with that of the $\Lambda_b^0 \rightarrow J/\psi \Lambda^0$ decay
has been made
with the ATLAS detector in proton--proton collisions at $\sqrt{s}=8\,$TeV at the LHC
using an integrated luminosity of $20.6\,$fb$^{-1}$.
The $J/\psi$ and $\psi(2S)$ mesons are reconstructed in their
decays to a muon pair, while the $\Lambda^0\rightarrow p\pi^-$ decay is
exploited for the $\Lambda^0$ baryon reconstruction.
The $\Lambda_b^0$ baryons are reconstructed
with transverse momentum $p_{\rm T}>10\,$GeV and pseudorapidity
$|\eta|<2.1$. 
The measured branching ratio of the
$\Lambda_b^0 \rightarrow \psi(2S) \Lambda^0$ and
$\Lambda_b^0 \rightarrow J/\psi \Lambda^0$ decays is
$\Gamma(\Lambda_b^0 \rightarrow \psi(2S)\Lambda^0)/\Gamma(\Lambda_b^0 \rightarrow J/\psi\Lambda^0) = 0.501\pm 0.033 ({\rm stat})\pm 0.019({\rm syst})$,
lower than the expectation from the covariant quark model.
}

\AtlasCoverSupportingNote{Supporting note}{https://cds.cern.ch/record/1666889}
\AtlasCoverCommentsDeadline{XX 07 2015}

\AtlasCoverAnalysisTeam{Alastair~Dewhurst, Leonid~Gladilin, Vladimir~Lyubushkin, Daniel~Scheirich}
\AtlasCoverEdBoardMember{Andreas~Korn~(chair)}
\AtlasCoverEdBoardMember{Evelina~Vassileva~Bouhova-Thacker}
\AtlasCoverEdBoardMember{Homer~Neal}
\AtlasCoverEdBoardMember{Irene~Vichou}

\AtlasCoverEgroupEditors{atlas-BPHY-2013-08-editors@cern.ch}

\AtlasCoverEgroupEdBoard{atlas-BPHY-2013-08-editorial-board@cern.ch}

\usepackage{cite}

\newcommand{\lb}  {\mbox{$\Lambda_b^0$}}
\newcommand{\lz}  {\mbox{$\Lambda^0$}}
\newcommand{\prob}{\mbox{$\cal P$}}
\newcommand{\acc} {\mbox{$\cal A$}}
\newcommand{\bran}{\mbox{$\cal B$}}

\def\lbjpsil0{ \Lambda_b^0 \rightarrow J/\psi \Lambda^0 }
\def\lbpsi2l0{ \Lambda_b^0 \rightarrow \psi(2S) \Lambda^0 }
\def\bjpsik0s{ B^0 \rightarrow J/\psi K^0_S }
\def\bpsi2k0s{ B^0 \rightarrow \psi(2S) K^0_S }

\begin{document}

\maketitle

\tableofcontents

\section{\label{secInt}Introduction}

The $\lb$ baryon properties have been extensively
studied at the Large Hadron Collider
(LHC)~\cite{ATLASlbtime,ATLASlbpol,CMSlbtime,CMSlbxsec,LHCblbmass,LHCblbtime,LHCblbpol}.
The decay channel $\lb \rightarrow J/\psi(\mu^+\mu^-) \Lambda^0(p\pi^-)$\footnote
{Hereafter, charge conjugation is implied,
unless explicitly stated otherwise.}
has been primarily used by the LHC experiments in these studies,
although a number of other $\lb$ decay channels have been exploited by the LHCb
experiment.
In particular, a measurement of the differential branching fraction
and angular analysis
of the rare decay $\lb \rightarrow \mu^+\mu^- \Lambda^0$ was performed
by LHCb~\cite{LHCblb2mul,LHCblb2mul2} following
observation of this decay by the CDF experiment~\cite{CDFlb2mul}
at the Tevatron collider.
However, no results for the decay mode $\lb \rightarrow \psi(2S) \Lambda^0$ have yet been reported,
although a measurement of the decay properties
would be useful for verification of theoretical predictions~\cite{gutsche}.

The $\lb \rightarrow \psi(2S) \Lambda^0$ branching fraction
should be of the same order as
that of the decay $\lb \rightarrow J/\psi \Lambda^0$
as suggested by the branching fraction values of the
$B^0$, $B^+$ and $B^0_s$ meson decays to $\psi(2S)/J/\psi$ and
either a pseudoscalar ($K^0$, $K^+$, $\eta$) or vector ($K^{*0}$, $K^{*+}$, $\phi$) meson.
The branching ratios
of such $B$ meson decays to $\psi(2S) X$ and $J/\psi X$ are
within the 0.5--0.8 range~\cite{pdg2014},
and are generally reproduced by factorisation calculations~\cite{kamal}.
The only available theoretical calculation of the branching ratio
of the
$\lb \rightarrow \psi(2S) \Lambda^0$ and
$\lb \rightarrow J/\psi \Lambda^0$ decays,
performed in the framework of the covariant quark model~\cite{model},
predicts $0.8$ with an uncertainty of approximately $0.1$~\cite{gutsche}.

An observation of the
$\lb \rightarrow \psi(2S) \Lambda^0$ decay and a measurement of 
the branching ratio
of the
$\lb \rightarrow \psi(2S) \Lambda^0$ and
$\lb \rightarrow J/\psi \Lambda^0$ decays
is reported in this Letter.
The $J/\psi$ and $\psi(2S)$ mesons are reconstructed in their
decays to a muon pair, while the $\Lambda^0\rightarrow p\pi^-$ decay is
exploited for the $\Lambda^0$ baryon reconstruction.
The $\Lambda_b^0$ baryons are reconstructed
with transverse momentum $p_{\rm T}>10\,$GeV and pseudorapidity
$|\eta|<2.1$. 

\section{\label{secDet}The ATLAS detector, data and Monte Carlo simulation samples}
A detailed description of the ATLAS detector can be found
elsewhere~\cite{ATLASPerformanceBook}. 
A brief outline of the components most relevant to this analysis
is given below.

The ATLAS inner detector (ID) has full coverage\footnote
{The ATLAS coordinate system is a Cartesian right-handed system,
with the coordinate origin at the nominal interaction point.
The anti-clockwise beam direction defines the positive $z$-axis,
with the $x$-axis pointing to the centre of the LHC ring.
Polar ($\theta$) and azimuthal ($\phi$) angles are measured
with respect to this reference system.
The pseudorapidity is defined as $\eta=-\ln\tan(\theta/2)$.} in $\phi$, covers
the pseudorapidity range $|\eta|<2.5$
and operates inside an axial magnetic field of $2\,$T.
It consists of a silicon pixel
detector (Pixel),
a silicon microstrip detector (semiconductor tracker, SCT) and
a transition radiation tracker (TRT).
The inner-detector barrel (end-cap) parts consist of
3 ($2\times3$) Pixel layers,
4 ($2\times9$) double-layers of single-sided SCT strips
and 73 ($2\times160$) layers of TRT straws.
The ATLAS muon spectrometer (MS) covers
the pseudorapidity range $|\eta|<2.7$.
It consists of precision tracking chambers, fast trigger detectors
and a large toroidal magnet system generating an average
field of
$0.5\,$T in the barrel region ($|\eta|<1.05$) and
$1\,$T in the end-cap regions ($1.05<|\eta|<2.7$).

The ATLAS detector has a three-level trigger system~\cite{ATLAStrigger}:
the hardware-based Level-1 system and the two-stage High Level Trigger (HLT).
For this measurement, dimuon triggers are used.
At Level-1, the dimuon triggers search for patterns of MS hits
corresponding to dimuons passing various $\pt$ thresholds.
Since the rate from the low-$\pt$ dimuon triggers was too high,
prescale factors were applied to reduce their output rates.
The data sample used in this analysis was collected
using three dimuon triggers with $\pt$ thresholds
of $4\gev$ for both muons, $4\gev$ and $6\gev$ for the two muons,
and $6\gev$ for both muons.
At the HLT, the dimuon triggers used require muons
with opposite charges and dimuon mass in the range
$2.5<m(\mu^+\mu^-)<4.3\,$GeV.

This analysis uses $20.6\,$fb$^{-1}$ of proton--proton
collision data with a centre-of-mass energy of $8\,$TeV recorded
by the ATLAS detector at the LHC in 2012.
The uncertainty on the integrated luminosity is $\pm2.8\%$.
It is derived following the same methodology as that
detailed in~\cite{ATLASlumi12}.
The event sample is processed using the standard offline ATLAS detector
calibration and event reconstruction code.
There are typically a few primary vertex candidates in each event
due to multiple collisions per bunch crossing.
Only events with
at least four reconstructed tracks with $p_{\rm T}>0.4\,$GeV
and at least one reconstructed primary vertex candidate
are kept for further analysis.

To model inelastic $pp$ events containing
$\lb \rightarrow J/\psi(\mu^+\mu^-) \Lambda^0$,
$\lb \rightarrow \psi(\mu^+\mu^-) \Lambda^0$,
$B^0 \rightarrow J/\psi(\mu^+\mu^-) K^0_S$ or
$B^0 \rightarrow \psi(\mu^+\mu^-) K^0_S$ decays,\footnote
{In this Letter, $\psi(2S)$ is referred to as $\psi$ when its decay channel is indicated.}
four large samples of Monte Carlo (MC) simulated events
are prepared using
the PYTHIA~8.1~\cite{pythia8} MC generator.
The $B^0$ MC samples are needed to control reflections from $B^0$ decays
to the $\lb$ signal distributions. 
The generation is based on leading-order matrix elements
for all $2\rightarrow 2$ QCD processes.
Initial- and final-state parton showering is used
to simulate higher-order processes.
Generated events
with both muons from $J/\psi$ or $\psi(2S)$ decays having transverse momenta above $3.5\,$GeV
and pseudorapidities within $\pm 2.5$, and, for $\lb$ MC samples, with the $\lz$ transverse momentum
above $1\,$GeV
are passed through a full simulation of the detector
using the ATLAS simulation framework~\cite{ATLASGEANT4}
based on GEANT4~\cite{GEANT4a, GEANT4b}
and processed with the same reconstruction program as used for the data.
An emulation of the three triggers used for the data collection
is applied to the MC samples.
The angular decay distributions of the
$\lb \rightarrow J/\psi(\mu^+\mu^-) \Lambda^0 (p\pi^-)$ decay
are modelled using the helicity amplitudes measured by ATLAS~\cite{ATLASlbpol}.
For the $\lb \rightarrow \psi(\mu^+\mu^-) \Lambda^0 (p\pi^-)$ decay,
the helicity amplitudes are set
to the predicted values~\cite{gutsche}.

\section{\label{secSel}Event and $\lb$ candidate selection}

\subsection{\label{secSel}Charmonium candidate selection}

Events are required to contain at least two muons identified
by the MS with tracks reconstructed in the ID.
The reconstructed muons are required to match
the muon candidates identified by the trigger.
The muon track parameters are taken from the ID measurement alone,
since the MS does not significantly improve the precision
in the momentum range
relevant for the charmonium measurements
presented here.
To ensure accurate measurements, each muon track must contain
at least six SCT hits and at least one Pixel hit.
Muon candidates satisfying these criteria are required to have
opposite charges
and a successful fit to a common vertex with $\chi^2/N_{\rm dof}<10$,
where $\chi^2$ is the fit quality
with the number of degrees of freedom $N_{\rm dof}=1$.
Events with $m(\mu^+\mu^-)$ values within
$\pm 200\,$MeV intervals around the $J/\psi$ and $\psi(2S)$ world average
masses~\cite{pdg2014} are
used to search for
$\Lambda^0\rightarrow p\pi^-$ candidates.

\subsection{\label{secSel}$\lz$ and $\bar{\Lambda}^0$ candidate selection}

In all events with $J/\psi$ or $\psi(2S)$ candidates,
pairs of tracks from particles with
opposite charge
are combined to form $\Lambda^0$ candidates.
Each track is required to have at least one Pixel or SCT hit.
Only pairs successfully fitted
to a common vertex with $\chi^2/N_{\rm dof}<5$ are kept.
The track with larger momentum is assigned the proton mass hypothesis
since the proton always has a larger momentum than the pion
for $\lz$ baryons with momenta larger than $0.3\,$GeV.
To suppress combinatorial background
the following requirements are used:
\begin{itemize}
\item{
$\pt(p)>1.7\,$GeV.
}
\item{
$|z_0(p)|<25\,{\rm mm}$,
where $z_0(p)$ is the proton longitudinal impact parameter with respect to the dimuon vertex.
MC studies show the requirement produces no loss of signal.
}
\item{
$L_{xy}^{\rm BL}(\Lambda^0)>7\,$mm, where $L_{xy}^{\rm BL}(\Lambda^0)$
is the transverse decay length\footnote
{The transverse decay length of a particle is the transverse distance between
the primary or production vertex and the particle decay vertex projected along
the transverse momentum of the particle.}
of the $\Lambda^0$ candidate
measured from the beam line.
}
\end{itemize}

Events with $m(p\pi^-)$ values within
an interval of $\pm 20\,$MeV around the $\Lambda^0$ world average
mass~\cite{pdg2014} are kept for further analysis.

\begin{figure}[]
  \includegraphics[width=0.5\textwidth]{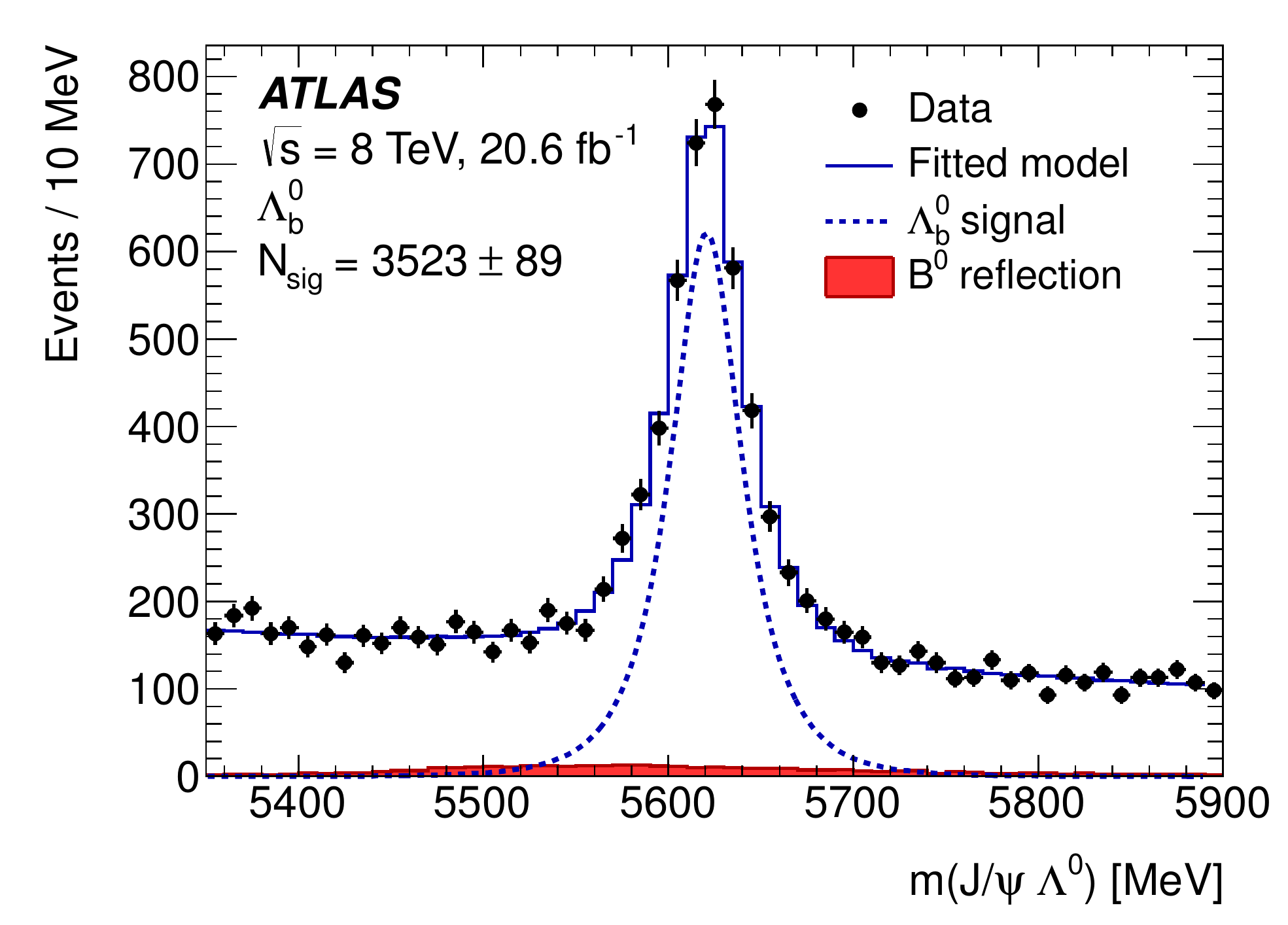}
  \includegraphics[width=0.5\textwidth]{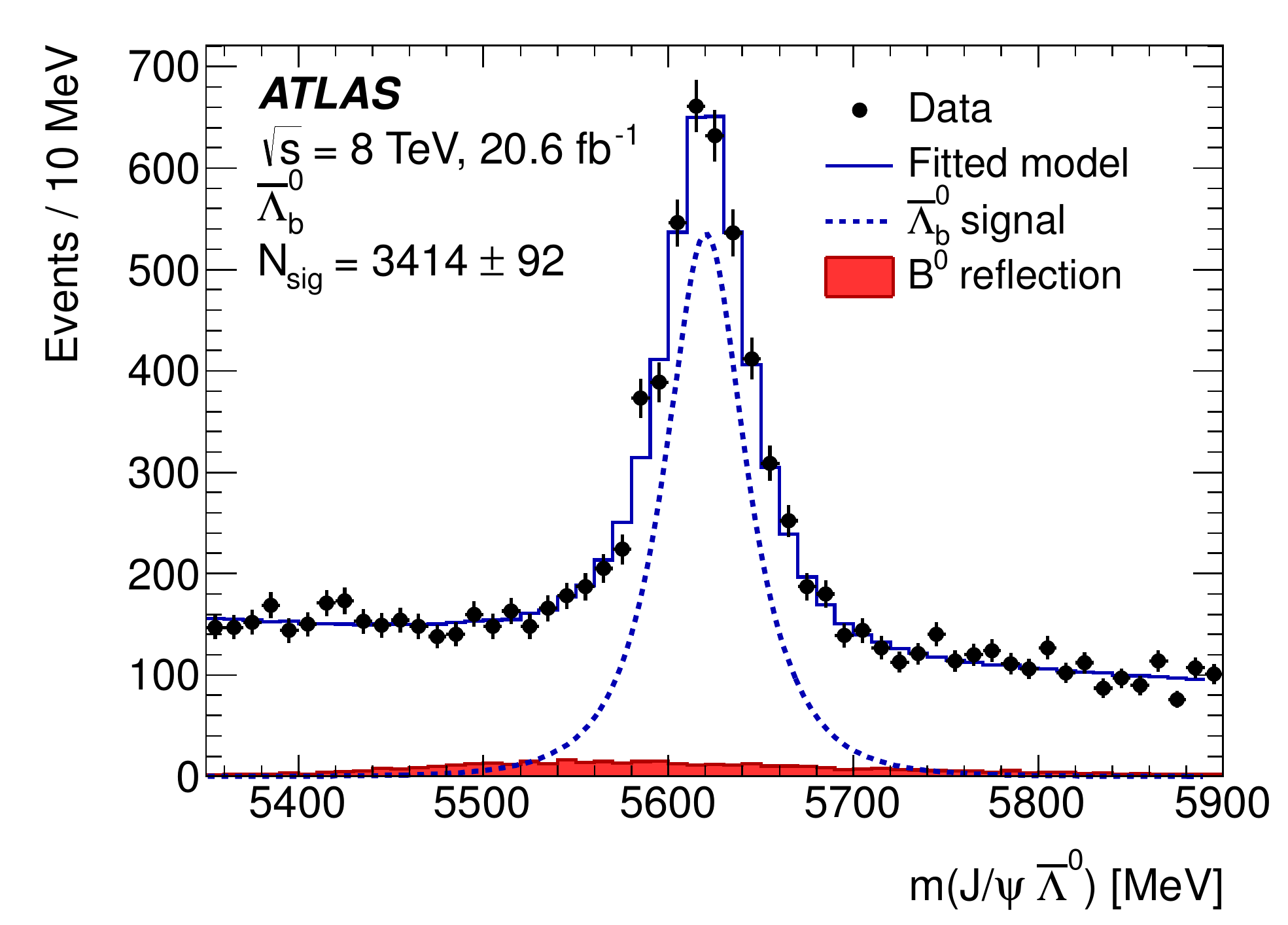}
  \caption{
The invariant mass distributions
$m(J/\psi\Lambda^0)$ (left plot)
and $m(J/\psi {\bar\Lambda}^0)$ (right plot)
for selected
$\Lambda_b^0$ and ${\bar\Lambda}_b^0$
candidates, respectively.
The solid histograms represent the fit results (see text).
The $\Lambda_b^0$ signals (dashed lines) and the $B^0$
reflections are also shown.
\label{fig:lambjpsi3b}}
\end{figure}

\begin{figure}[]
  \includegraphics[width=0.5\textwidth]{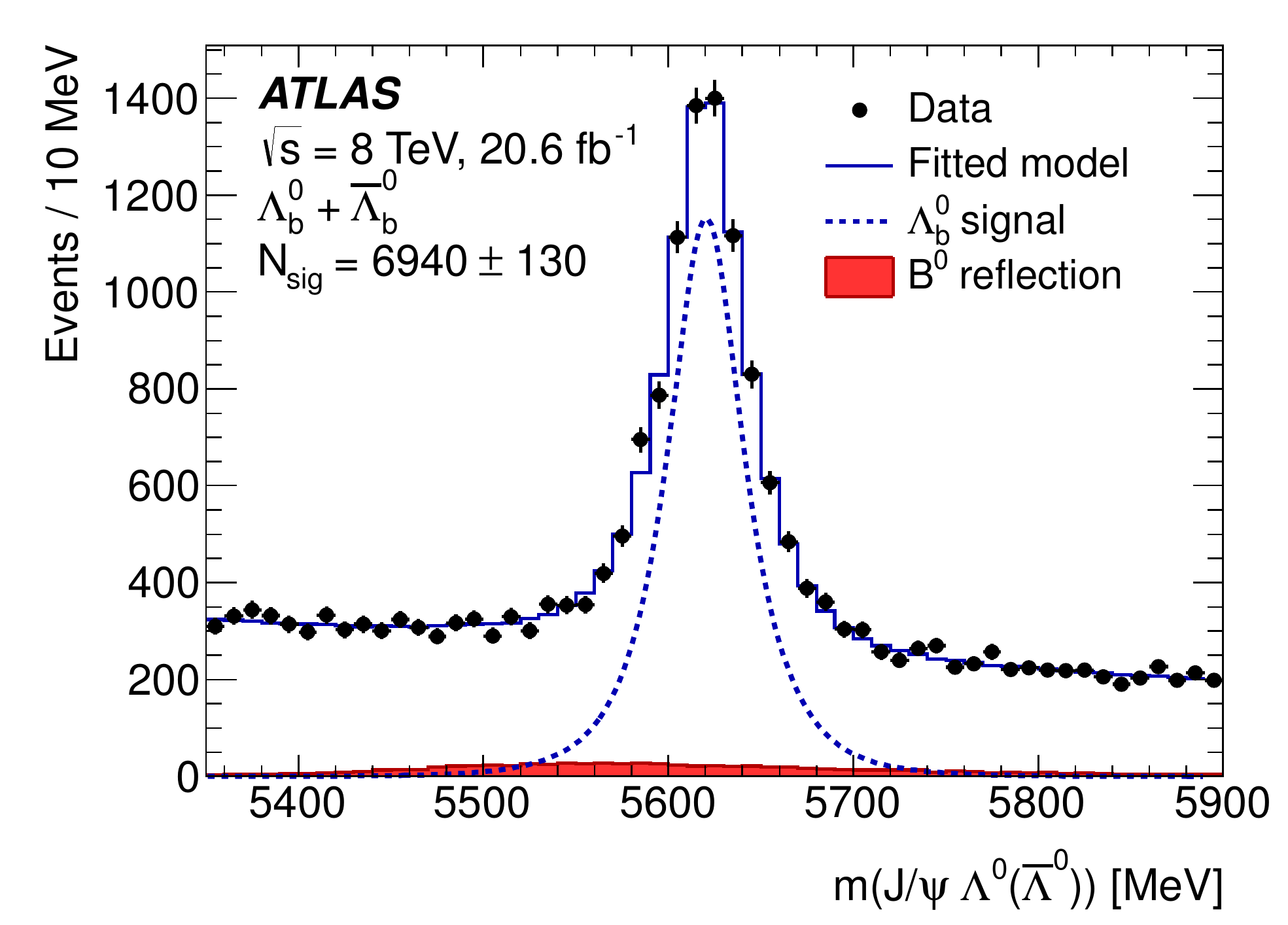}
  \includegraphics[width=0.5\textwidth]{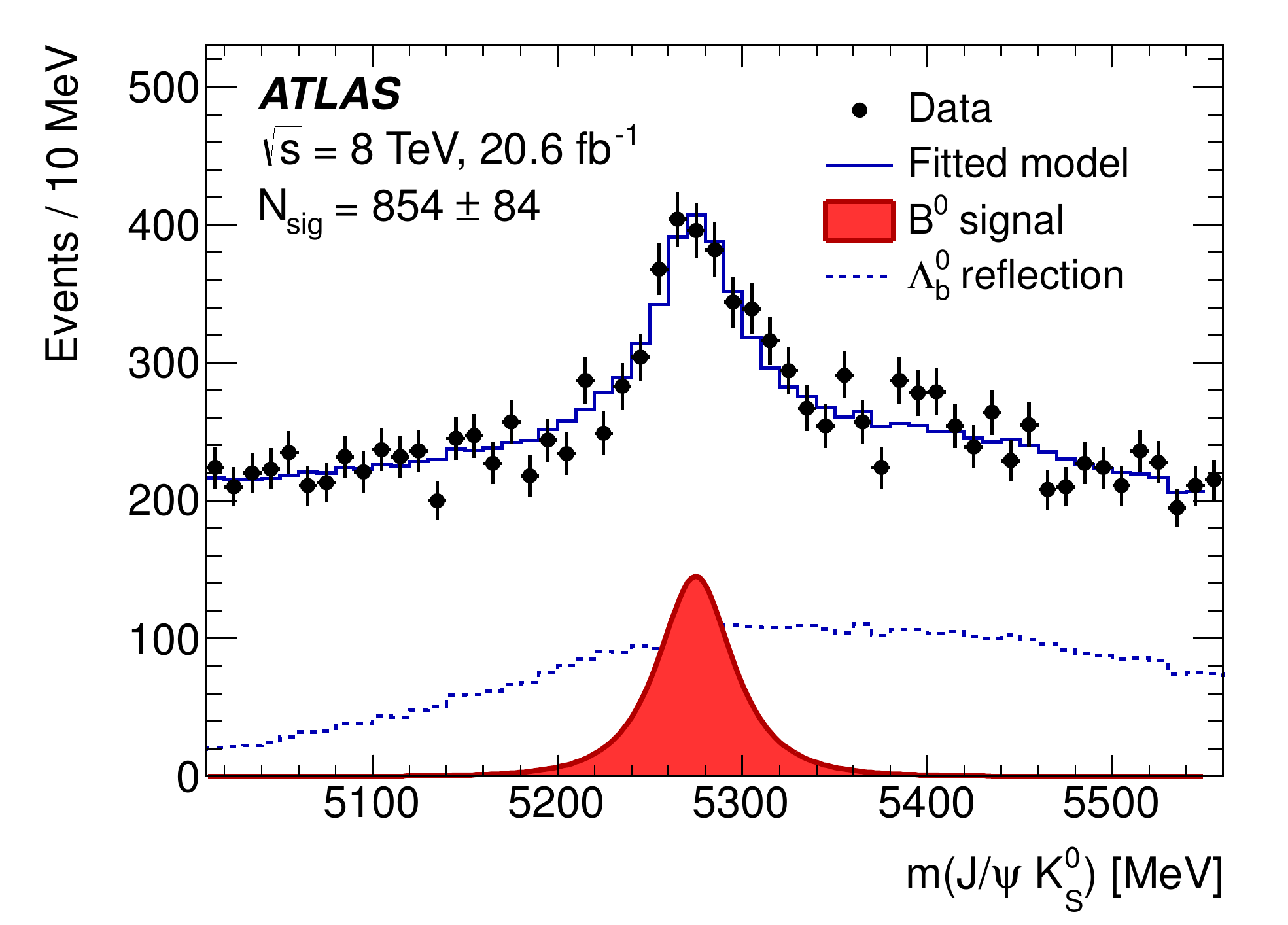}
  \caption{
The invariant mass distributions for the combined sample of the selected
$\Lambda_b^0$ and
${\bar\Lambda}_b^0$
candidates
obtained after their fits to the
$\lb \rightarrow J/\psi \Lambda^0$ (left plot)
and
$B^0 \rightarrow J/\psi K^0_S$ (right plot)
topologies.
The solid histograms represent fit results (see text).
The $\Lambda_b^0$ and $B^0$ signals and their mutual
reflections are also shown.
\label{fig:lambjpsi3}}
\end{figure}

\begin{figure}[]
  \includegraphics[width=0.5\textwidth]{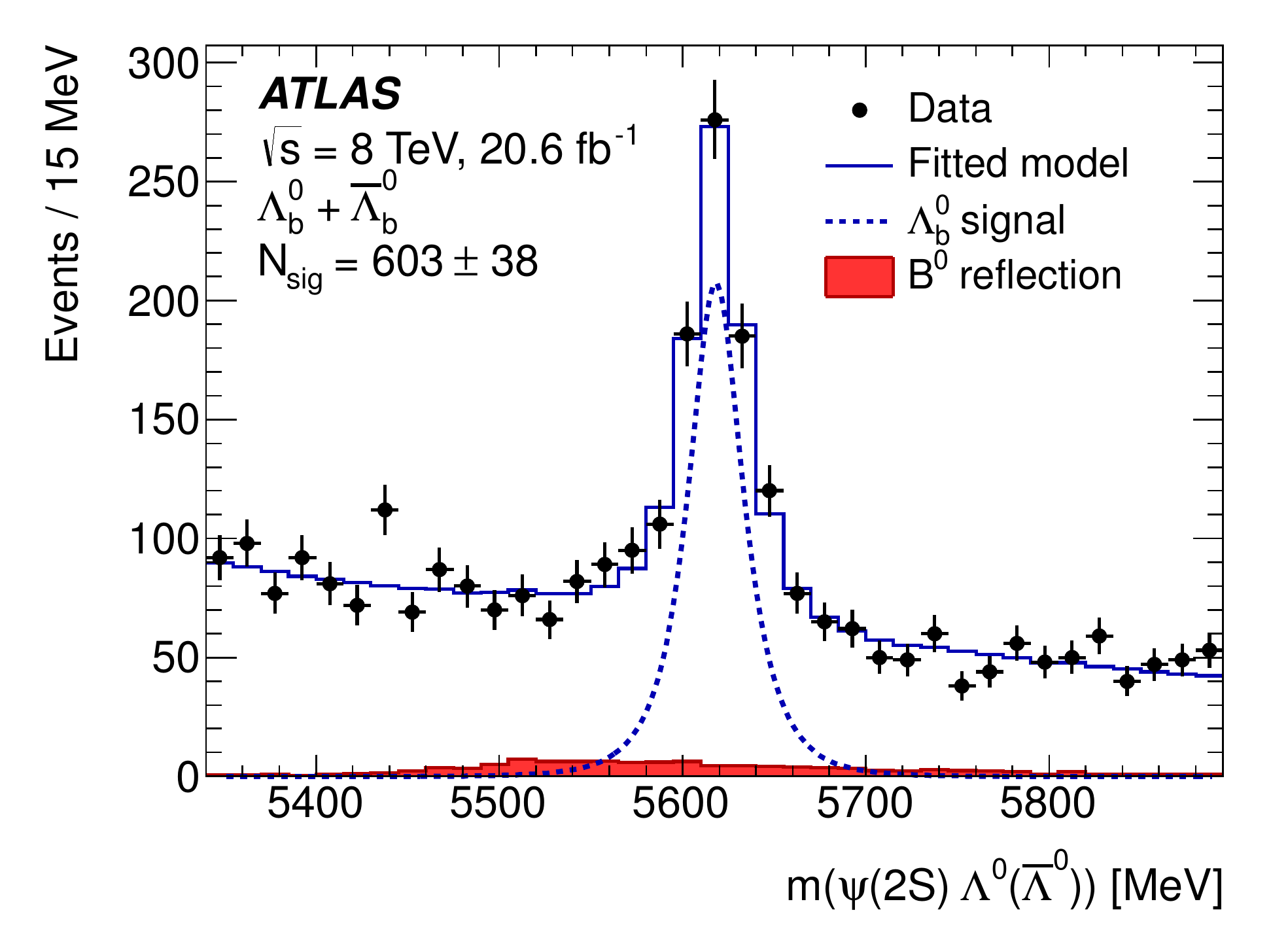}
  \includegraphics[width=0.5\textwidth]{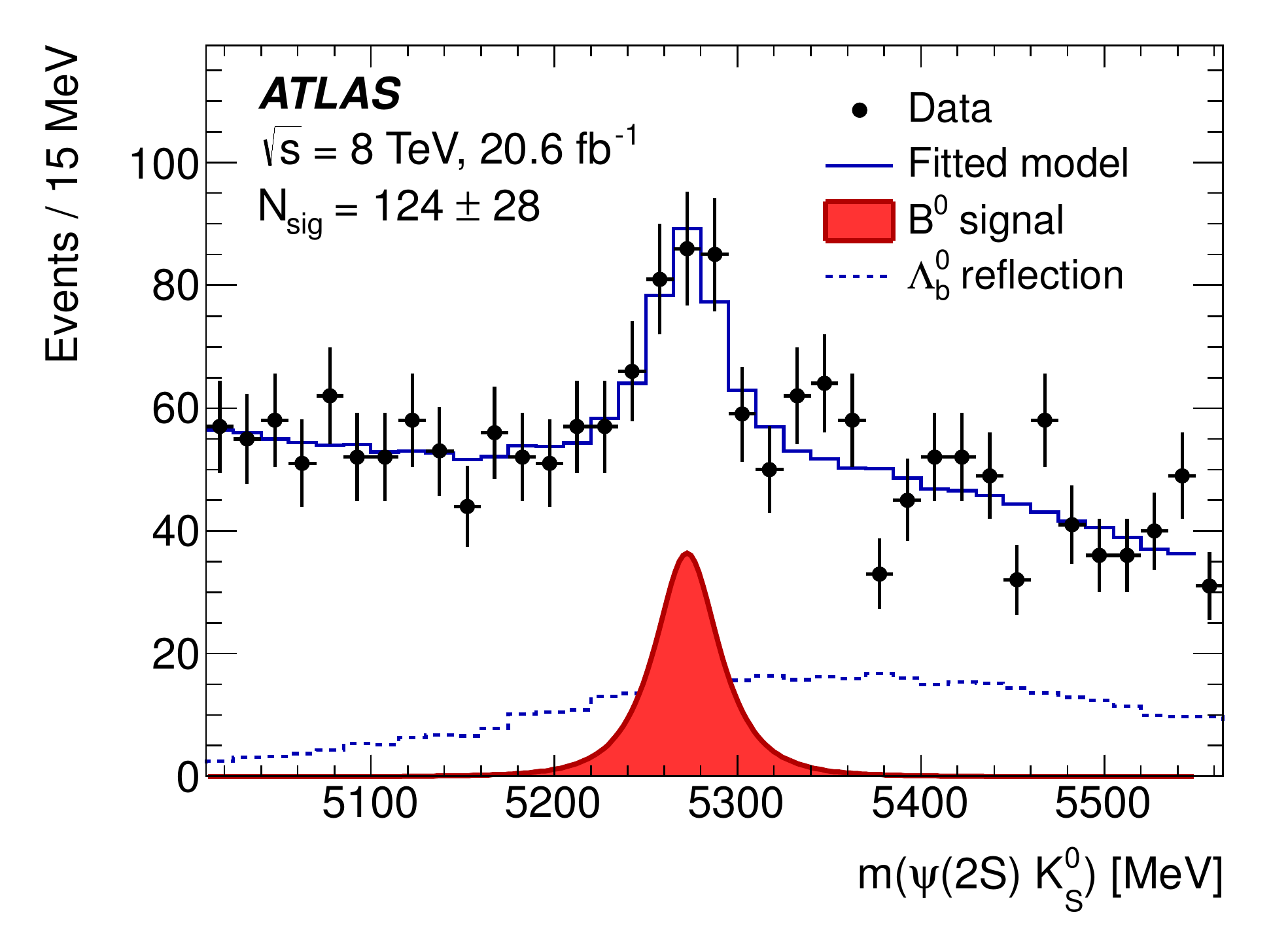}
  \caption{
The invariant mass distributions for the combined sample of the selected
$\Lambda_b^0$ and
${\bar\Lambda}_b^0$
candidates
obtained after their fits to the
$\lb \rightarrow \psi(2S) \Lambda^0$ (left plot)
and
$B^0 \rightarrow \psi(2S) K^0_S$ (right plot)
topologies.
The solid histograms represent fit results (see text).
The $\Lambda_b^0$ and $B^0$ signals and their mutual
reflections are also shown.
\label{fig:lambpsi2s3}}
\end{figure}

\subsection{\label{secSel}$\lb$ reconstruction}

Tracks of the selected charmonium and $\lz$ candidates
are simultaneously refitted with the
dimuon and dihadron masses constrained to the world average masses
of $J/\psi$ ($m_{J/\psi}$) or $\psi(2S)$ ($m_{\psi(2S)}$)
and $\lz$ ($m_{\Lambda^0}$)~\cite{pdg2014}, respectively.
The combined momentum of the refitted $\lz$ track pair
is required to point to the dimuon vertex. 
To control $B^0$ reflections to the $\lb$ signal distributions,
a $B^0$ decay topology fit is also attempted
for each track quadruplet successfully fitted to the
$\lb$ topology,
i.e. the pion mass is assigned to both hadron tracks
and the dihadron mass is constrained to the world average
mass of $K_S^0$~\cite{pdg2014}.
To suppress combinatorial and $B^0$ backgrounds
the following requirements are used:
\begin{itemize}
\item{
$\chi^2(\Lambda_b^0)/N_{\rm dof}<3$,
where $\chi^2$ is the quality
of the fit to the $\lb$ topology
with $N_{\rm dof}=6$.
}
\item{
$L_{xy}(\Lambda^0)>10\,$mm, where $L_{xy}(\Lambda^0)$
is the transverse decay length of the refitted $\Lambda^0$ vertex
measured from the $\lb$ (dimuon) vertex.
}
\item{
$\pt(\Lambda^0)>2.5\,$GeV.
}
\item{
$\pt(\pi^-)>0.45\,$GeV.
}
\item{
$\tau(\Lambda_b^0)>0.35\,$ps,
where
$\tau(\Lambda_b^0)=L_{xy}(\Lambda_b^0)\cdot m_{\Lambda_b^0}/p_{\rm T}(\Lambda_b^0)$
is the $\lb$ proper decay time,
$L_{xy}(\Lambda_b^0)$ is
the transverse decay length
of the $\lb$ vertex
measured from the primary vertex
and $m_{\Lambda_b^0}$ is
the $\lb$ world average mass~\cite{pdg2014}.
The primary vertex candidate with
at least three tracks
and the smallest value of
the three-dimensional impact parameter 
of the $\Lambda_b^0$ candidate
is selected as the actual primary vertex.}
\item{
$\prob(\Lambda_b^0)>\prob(B^0)$,
where $\prob(\Lambda_b^0)$ and $\prob(B^0)$
are the $\chi^2$ probabilities of the quadruplet fits
with $\lb$ and $B^0$ topologies, respectively.
}
\end{itemize}
The muon transverse momenta and pseudorapidities
are required to be in the ranges with high values of the trigger
and reconstruction acceptances:

$$\pt(\mu^\pm)>4\,{\rm GeV,}~|\eta(\mu^\pm)|<2.3.$$
The kinematic range of the $\lb$ measurement is fixed
to

$$\pt(\lb)>10\,{\rm GeV,}~|\eta(\lb)|<2.1.$$

The invariant mass distribution $m(J/\psi\lz)$,
calculated using track parameters from the $\lb$ topology fits,
is shown in
Figure~\ref{fig:lambjpsi3b}
separately
for the selected
$\Lambda_b^0$
and ${\bar\Lambda}_b^0$
candidates.
Clear signals with similar size
are seen in the two distributions
around the world average mass of the $\lb$ baryon.
Figures~\ref{fig:lambjpsi3} and~\ref{fig:lambpsi2s3}
show the $m(J/\psi\lz)$ and $m(\psi(2S)\lz)$
distributions for the combined sample of the $\Lambda_b^0$ and 
${\bar\Lambda}_b^0$ candidates.
The invariant mass distributions $m(J/\psi K^0_S)$ and $m(\psi(2S) K^0_S)$
from the $B^0$ topology fits
are also shown.
Clear signals are seen
in the $m(J/\psi\lz)$ and $m(\psi(2S)\lz)$ distributions\footnote{
Studies with MC simulated events show that the fraction of reconstructed $\lbjpsil0$ decays
which can contribute to
the reconstructed $\lbpsi2l0$ signal is $\sim10^{-5}$.
}
around the world average mass of the $\lb$ baryon.
There are also signals
in the $m(J/\psi K^0_S)$ and $m(\psi(2S) K^0_S)$ distributions
near the world average mass of the $B^0$ meson~\cite{pdg2014}.
The $B^0$ signals are smaller than the $\Lambda_b^0$ signals
due to the selection requirements.

The $m(J/\psi\lz)$ and $m(J/\psi K^0_S)$ distributions are simultaneously
fitted to sums of signal and two-component background distributions.
The signals are described
by modified Gaussian functions~\cite{ZEUSJpsi}.
The modified Gaussian function is defined as

$${\rm Gauss}^{\rm mod}\propto \exp [-0.5 \cdot x^{1+1/(1+0.5 \cdot x)}],$$
where $x=|(m-m_0)/\sigma|$.
This functional form,
introduced to take into account the non-Gaussian tails of resonant signals,
describes both data and MC signals well.
The signal position, $m_0$,
and width, $\sigma$, as well as the number of the signal events
are free parameters of the fit.
The non-resonant backgrounds in the distributions are described
by independent exponential functions.
The mutual $B^0$ and $\Lambda_b^0$ reflections are described by MC templates
normalised to the numbers of $B^0$ and $\lb$ hadrons obtained
in the fit.
The reflection normalisations are corrected for
small losses (2--6$\%$)
of $\lb$ and $B^0$ hadrons that passed the $\lb$ reconstruction
but failed the $B^0$ reconstruction.
The corrections are obtained using MC simulation.
A similar fit is performed for
the $m(\psi(2S)\lz)$ and $m(\psi(2S) K^0_S)$ distributions.
In the analysis of the combined $\Lambda_b^0$ and 
${\bar\Lambda}_b^0$ samples,
the ratio of the MC $\Lambda_b^0$ and ${\bar\Lambda}_b^0$ events
is set to the data ratio
obtained in the
separate
$\Lambda_b^0 \rightarrow J/\psi\lz$ and
$\bar{\Lambda}_b^0 \rightarrow J/\psi\bar{\Lambda}^0$
fits (Figure~\ref{fig:lambjpsi3b}).
The $\Lambda_b^0$ and ${\bar{\Lambda}}_b^0$ fitted yields are
$3523\pm 89$ and $3414\pm 92$, respectively, providing the ratio
$1.03\pm 0.04({\rm stat})$.

The results of the fits
for the combined $\Lambda_b^0$ and 
${\bar\Lambda}_b^0$ samples
are summarised in Table~\ref{tab:lbb0num}.
The $\lb$ mass values obtained from the fits of
the $m(J/\psi\lz)$ and $m(\psi(2S)\lz)$ distributions
agree with each other and with the world average $\lb$ mass value~\cite{pdg2014}.
The signal widths are different,
reflecting the difference in charmonium masses
in the two decay channels,
in agreement with the MC expectations.
The quality,
$\chi^2/N_{\rm dof}$, 
of the
$\lbjpsil0$ and $\lbpsi2l0$ signal fits
are $1.0$ and $1.1$, respectively.

\begin{table}[h!]
\begin{center}
\renewcommand{\arraystretch}{1.4}
\begin{tabular}{c|c|c|c|c} \hline
& $\lbjpsil0$ & $\bjpsik0s$ & $\lbpsi2l0$ &  $\bpsi2k0s$ \\
\hline
%\hline
 $N_{\rm sig}$ & $6940 \pm 130$ & $854 \pm 84$ & $603 \pm 38$ & $124 \pm 28$ \\
%\hline
 $m_{\rm sig}\,$[MeV] & $5620.4 \pm 0.4$ & $5274.7\pm2.3$ & $5618.2\pm 1.2$ & $5272.4\pm 4.9$ \\
%\hline
 $\sigma_{\rm sig}\,$[MeV] & $19.7\pm 0.5$ & $19.2\pm 2.2$ & $14.3\pm 1.1$ & $16.7\pm 4.1$ \\
\hline
\end{tabular}
\caption{
The numbers of signal events, $N_{\rm sig}$,
signal masses, $m_{\rm sig}$, and signal widths, $\sigma_{\rm sig}$,
obtained by the fits (see text).
Only statistical uncertainties are shown.
}
%\vspace*{0.5cm}
\label{tab:lbb0num}
\end{center}
\end{table}

\begin{figure}[]
  \includegraphics[width=0.5\textwidth]{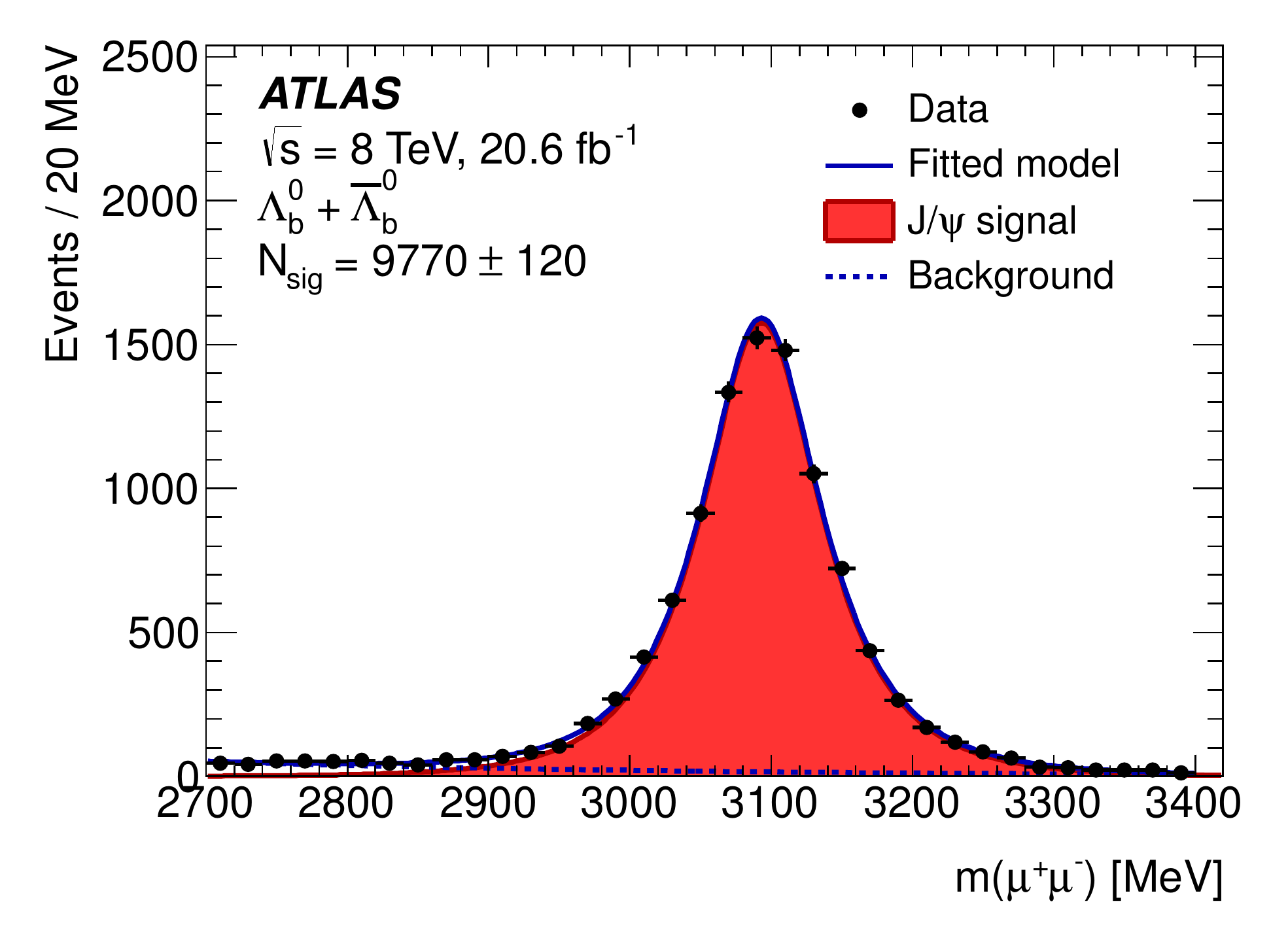}
  \includegraphics[width=0.5\textwidth]{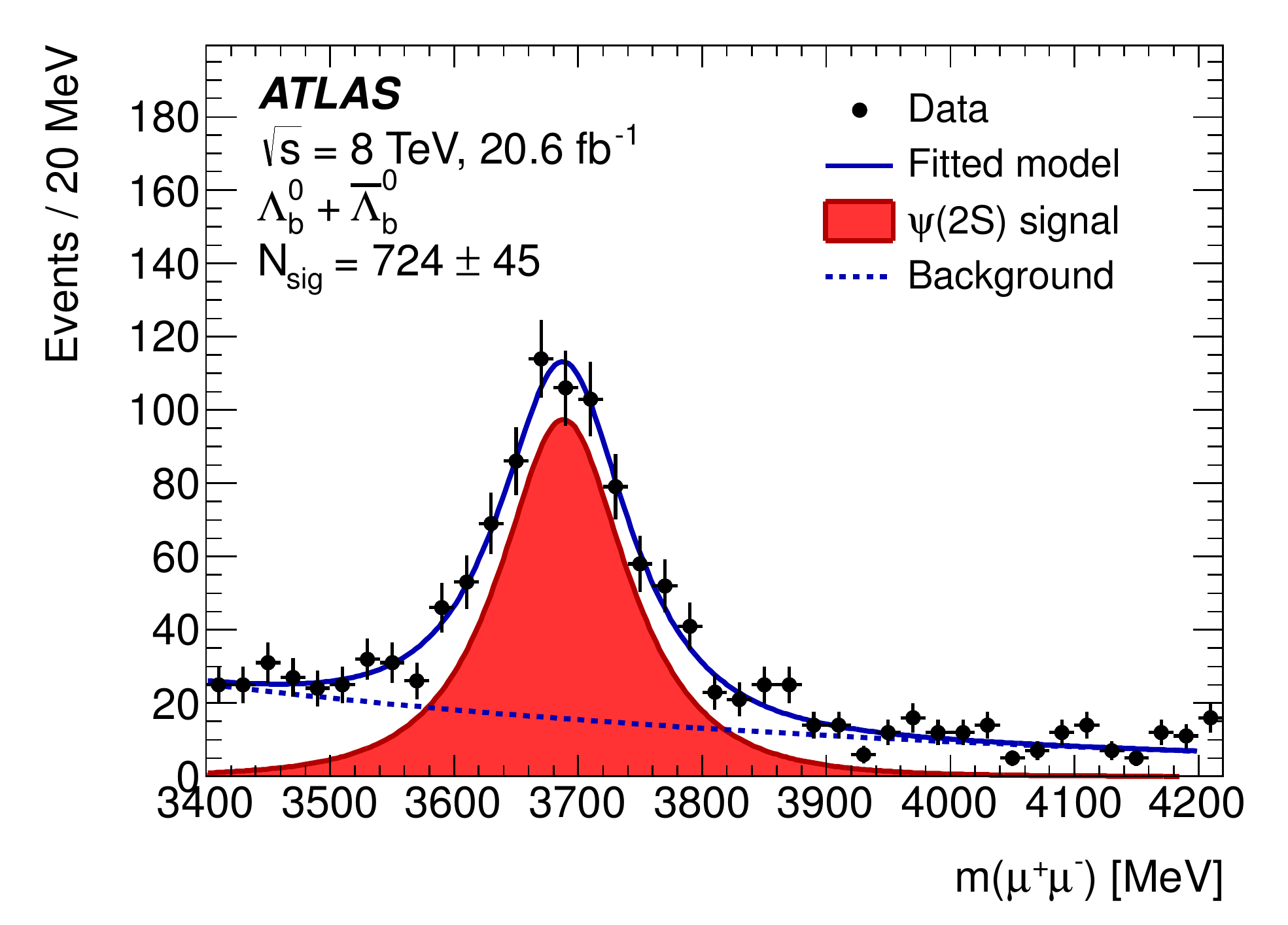}
  \caption{
The
$m(\mu^+\mu^-)$ distributions for
$\lbjpsil0$ candidates (left plot) and
$\lbpsi2l0$ candidates (right plot)
after full selection, without a mass constraint
on the charmonium mass
in the cascade fit.
The spectra are fitted with a sum of
an exponential function and a modified Gaussian function.
\label{fig:jpsipsi2s}}
\end{figure}

\begin{figure}[]
  \includegraphics[width=0.5\textwidth]{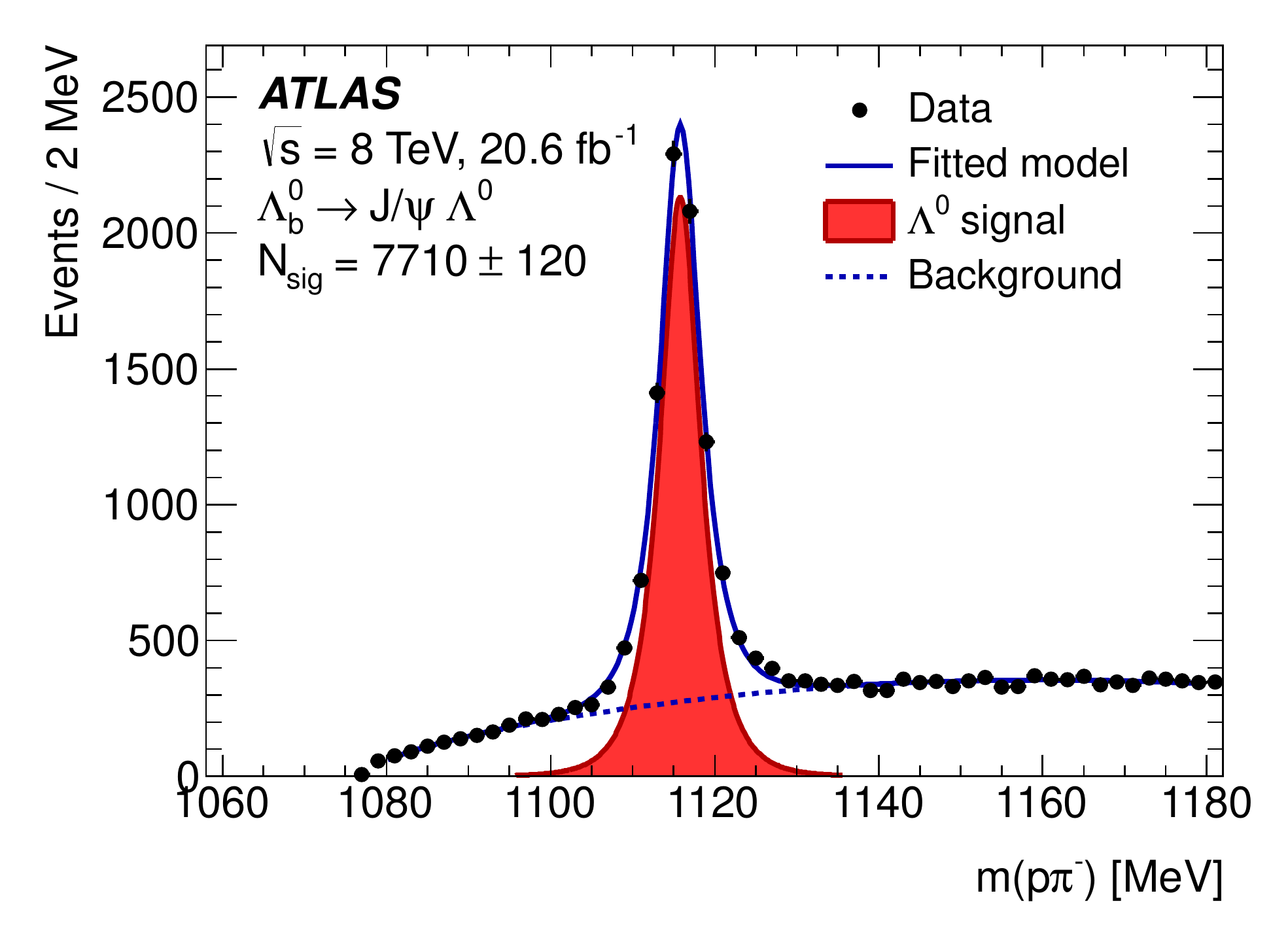}
  \includegraphics[width=0.5\textwidth]{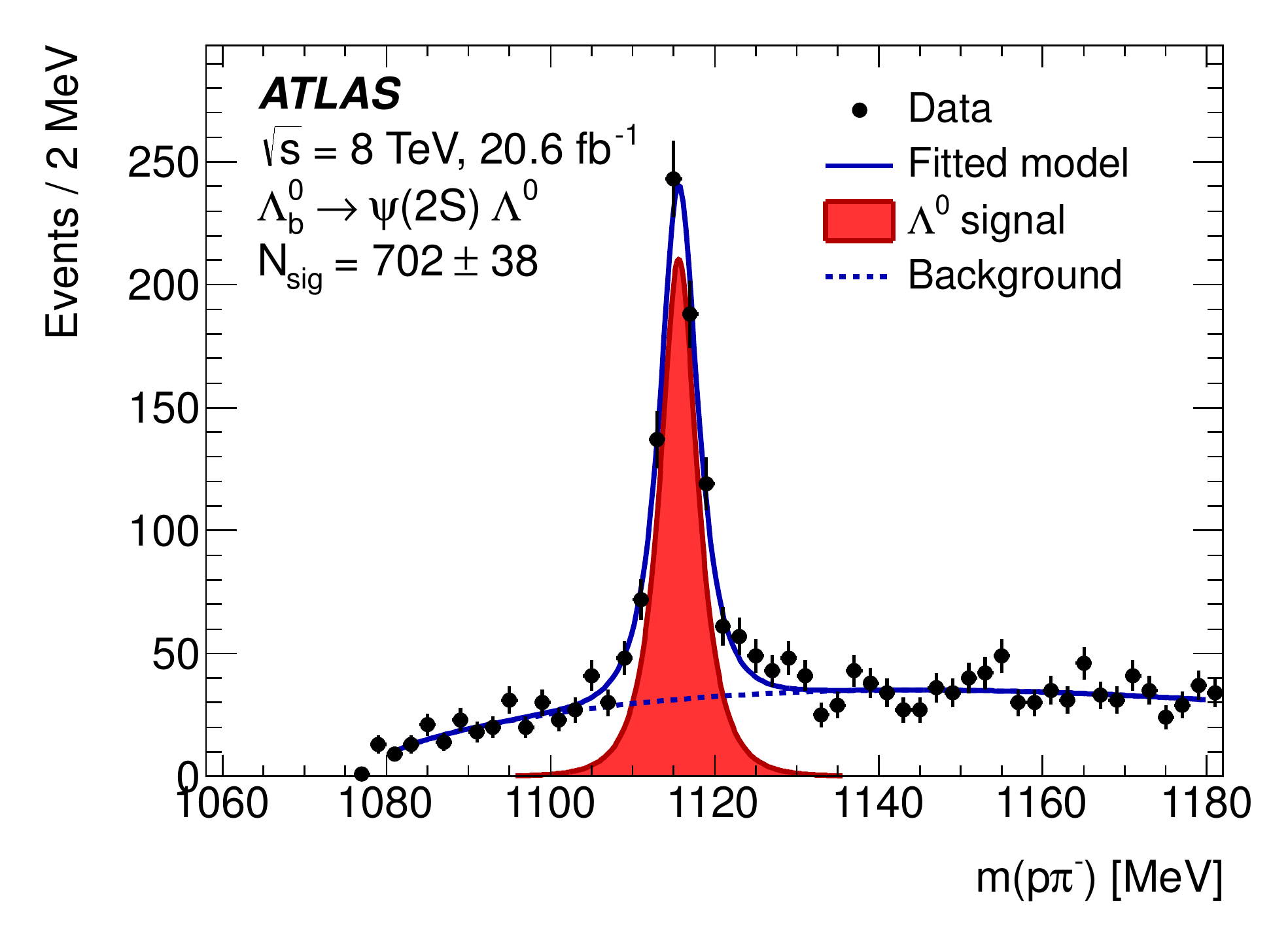}
  \caption{
The
$m(p\pi^-)$ distributions for
$\lbjpsil0$ candidates (left plot) and
$\lbpsi2l0$ candidates (right plot)
after full selection, without a mass constraint
on the $\Lambda^0$ mass
in the cascade fit.
The spectra are fitted with a sum of
a threshold function and a modified Gaussian function.
\label{fig:lambda02}}
\end{figure}

To verify that the observed $\lb$ signals correspond to
the $\lbjpsil0$ and $\lbpsi2l0$ decays
the signal reconstruction is repeated with only one mass constraint
for either the dimuon or the dihadron track pair in the cascade fit
and the $\lb$ mass is calculated using the mass-difference method.
In the case that the dihadron mass is fixed to the $\lz$ mass,
the $\lb$ mass is calculated as
$m(\mu^+\mu^-\Lambda^0)-m(\mu^+\mu^-)+m_{J/\psi}$
($m(\mu^+\mu^-\Lambda^0)-m(\mu^+\mu^-)+m_{\psi(2S)}$) for $m(\mu^+\mu^-)<3.4\gev$
($m(\mu^+\mu^-)>3.4\gev$).
When the dimuon mass is fixed to the $J/\psi$ ($\psi(2S)$) mass,
the $\lb$ mass is calculated as
$m(J/\psi p\pi^-)-m(p\pi^-)+m_{\Lambda^0}$
($m(\psi(2S) p\pi^-)-m(p\pi^-)+m_{\Lambda^0}$).
In both cases clean $\lb$ signals are reconstructed with numbers
of signal events compatible with those in Table~\ref{tab:lbb0num}. 
Figure~\ref{fig:jpsipsi2s}
shows the $m(\mu^+\mu^-)$ distributions for
the $\lbjpsil0$ and $\lbpsi2l0$ candidates
reconstructed with the mass constraint for the dihadron pair
and selected within $\pm 3\sigma_{\rm sig}$ around the world average $\lb$ mass.
Clear signals from $J/\psi$ and $\psi(2S)$ are seen.
The $m(\mu^+\mu^-)$ distributions are fitted to a sum of an exponential
function describing the background and a modified Gaussian function describing the signal. The signal yields are found to be
$N_{J/\psi}=9770\pm 120$ and $N_{\psi(2S)}=724\pm 45$.
Figure~\ref{fig:lambda02}
shows the $m(p\pi^-)$ distributions for
the $\lb$ candidates
reconstructed with the mass constraint for the dimuon pair
and selected within $\pm 3\sigma_{\rm sig}$ around the world average $\lb$ mass.
Clear signals from $\Lambda^0$ are seen.
The $m(p\pi^-)$ distributions are fitted to a sum of a threshold
function describing the background and a modified Gaussian function describing
the signal.
The threshold function has the form

$$A\cdot (m -m_p-m_{\pi^-})^B\cdot\exp[C\cdot(m - m_p - m_{\pi^-})+D\cdot(m - m_p - m_{\pi^-})^2],$$
where $m_p$ and $m_{\pi^-}$ are the proton and pion
masses, respectively,
and $A$, $B$, $C$ and $D$ are free parameters.
The $\Lambda^0$ signal yields are found to be
$7710\pm 120$ and $702\pm 38$ for
the $\lbjpsil0$ and $\lbpsi2l0$ candidates, respectively.
The numbers of signal charmonium and $\Lambda^0$ events are larger
than the numbers of the corresponding $\lb$ signal events because
the backgrounds are partly due to genuine
charmonium and $\Lambda^0$ states. 

\section{\label{secMes}Measurement of the $\lb$ branching ratio $\Gamma(\lb \rightarrow \psi(2S)\Lambda^0)/\Gamma(\lb \rightarrow J/\psi\Lambda^0)$}

The numbers of $\lb$ signal events in the two decay modes,
reported in Table~\ref{tab:lbb0num},
are corrected for detector effects and selection efficiencies
as $N_{\rm cor} = N_{\rm sig} / \acc$,
where $N_{\rm cor}$ is the corrected number
and $\acc$ is the MC acceptance.
The MC events
with the $\psi(2S)/J/\psi$ muons having transverse momenta above $3.5\,$GeV
and pseudorapidities within $\pm 2.5$, and $\lz$ transverse momentum
above $1\,$GeV,
passed through the detector simulation and event reconstruction,
are used
to correct the numbers of signal events
in the fiducial range, defined as follows:

$$\pt(\lb)>10\,{\rm GeV,}~|\eta(\lb)|<2.1,$$
$$\pt(\mu^\pm)>4\,{\rm GeV,}~|\eta(\mu^\pm)|<2.3,$$
$$\pt(\lz)>2.5\,{\rm GeV}.$$
The acceptances are calculated as the ratio of the number of reconstructed
$\lb$ signal events
passing all selection requirements
in the above fiducial range
to the number of $\lb$ baryons in the same decay mode and fiducial range
at the MC generator level.   
These acceptances are
$4.16\pm0.02({\rm stat})\%$ and $4.30\pm0.03({\rm stat})\%$
for the
$\lb \rightarrow J/\psi\Lambda^0$ and $\lb \rightarrow \psi(2S)\Lambda^0$
decays, respectively.
In the fiducial range,
the ratio of the corrected numbers
of $\lb$ signal events in the two decay modes is
$0.0841\pm0.0055({\rm stat})$.

Then the numbers are corrected,
using generator-level MC samples
with no requirements on the $\mu^\pm$ and $\Lambda^0$
selection,
from the above fiducial range
to the kinematic range of the $\lb$ measurement

$$\pt(\lb)>10\,{\rm GeV,}~|\eta(\lb)|<2.1.$$
The acceptances of the latter corrections
are
$7.57\pm0.06({\rm stat})\%$ and $9.61\pm0.07({\rm stat})\%$
for the
$\lb \rightarrow J/\psi\Lambda^0$ and $\lb \rightarrow \psi(2S)\Lambda^0$
decays, respectively.
Finally, the branching ratio of the two $\lb$ decays
is calculated as

$$\frac{\Gamma(\lb \rightarrow \psi(2S)\Lambda^0)}{\Gamma(\lb \rightarrow J/\psi\Lambda^0)}=
\frac{N_{\rm cor}(\lb\rightarrow \psi(\mu^+\mu^-)\lz)}{N_{\rm cor}(\lb\rightarrow J/\psi(\mu^+\mu^-)\lz)}\cdot
\frac{\bran(J/\psi\rightarrow \ell^+\ell^-)}{\bran(\psi(2S) \rightarrow \ell^+\ell^-)},$$
where $\bran$ is the branching fraction of the corresponding charmonium decay to a lepton pair.
In the case of $J/\psi$, the branching fraction
$\bran(J/\psi\rightarrow \mu^+\mu^-) = 0.05961\pm0.00033$~\cite{pdg2014} is used.
For $\bran(\psi(2S) \rightarrow \ell^+\ell^-)$, the branching fraction
$\bran(\psi(2S)\rightarrow e^+e^-) = 0.00789\pm0.00017$ is used,
assuming lepton universality,
because it is
measured with better precision than
in the muon channel, 
$\bran(\psi(2S)\rightarrow \mu^+\mu^-) = 0.0079\pm0.0009$~\cite{pdg2014}.

Five groups of systematic uncertainty sources
are considered.
The effect of each group on the measured ratio,
obtained by adding in quadrature the effects of
independent sources,
is shown in parentheses:
\begin{itemize}
\item{
Dependence on the $\lb$ production model
($\pm0.1\%$).
The uncertainty is obtained by
\begin{itemize}
\item{varying
the MC $\pt(\lb)$ and
$|\eta(\lb)|$ distributions
while preserving agreement with the data distributions,}
\item{varying the MC ratio
of $\Lambda_b^0$ and ${\bar\Lambda}_b^0$ baryons
in the range allowed by the separate data fits (Section 3),}
\item{varying the lifetimes of the $\Lambda^0$ and $\Lambda_b^0$ baryons
in the ranges of their uncertainties~\cite{pdg2014}.}
\end{itemize}
}
\item{
Dependence on the $\lb$ polarisation model
($\pm 1.1\%$).
The uncertainty is obtained by varying
the MC $\lb \rightarrow J/\psi(\mu^+\mu^-) \Lambda^0 (p\pi^-)$ helicity amplitudes in the range of
their uncertainties~\cite{ATLASlbpol},
and by changing
the MC $\lb \rightarrow \psi(\mu^+\mu^-) \Lambda^0 (p\pi^-)$ helicity amplitudes
to those
measured by ATLAS for the $\lb \rightarrow J/\psi(\mu^+\mu^-) \Lambda^0 (p\pi^-)$
decay~\cite{ATLASlbpol}.
}
\item{
The uncertainty of the signal extraction procedures
($\pm2.8\%$).
The uncertainty is determined by
changing the background parameterisations
to second order polynomials
and by reducing
the ranges used for the signal fits
by $20\,$MeV from either left or right side,
independently for the two $\lb$ signals.
In addition, the corrections of the reflection normalisations,
obtained from MC simulation, are varied by half of their values.
This uncertainty is affected by statistical fluctuations.
}
\item{
The uncertainty originating from the MC statistical uncertainty
($\pm 1.3\%$).
}
\item{
The uncertainty of the charmonium branching fractions 
$\bran(J/\psi\rightarrow \mu^+\mu^-)$ and $\bran(\psi(2S)\rightarrow e^+e^-)$ ($\pm2.2\%$).
}
\end{itemize}

The measured branching ratio of the two $\lb$ decays
is

$$\frac{\Gamma(\lb \rightarrow \psi(2S)\Lambda^0)}{\Gamma(\lb \rightarrow J/\psi\,\Lambda^0)}=0.501\pm 0.033 ({\rm stat})\pm 0.016 ({\rm syst})\pm 0.011 ({\bran}),$$
where
the contributions from
the first four groups of systematic uncertainty
are added
in quadrature.
The uncertainty due to the uncertainties of the charmonium
branching fractions $\bran$ is quoted separately.
The luminosity uncertainty,
uncertainties of the muon and hadron track reconstruction
and the vertexing uncertainties cancel out in the ratio. 
The bias in the measured ratio due to contributions from
the rare decay $\lb \rightarrow \mu^+\mu^- \Lambda^0$
is estimated using the LHCb measurement~\cite{LHCblb2mul2}
of the rare decay's differential branching fraction
to be below $0.5\%$ and thus neglected.
Consistent ratio values are found when calculated
in bins of $p_{\rm T}(\Lambda_b^0)$ or
separately for $\lb$
and $\bar{\Lambda}_b^0$
baryons.

The measured ratio
lies in the range 0.5--0.8 found for the branching ratios
of analogous $B$ meson decays~\cite{pdg2014}. 
The only available calculation for the branching ratio of the two $\lb$ decays
($0.8\pm 0.1$~\cite{gutsche}) exceeds the measured value.

\section{\label{secSum}Summary}

The $\lb \rightarrow \psi(2S) \Lambda^0$ decay
has been observed
with the ATLAS detector in $pp$ collisions at $\sqrt{s}=8\,$TeV at the LHC
using an integrated luminosity of $20.6\,$fb$^{-1}$.
The branching ratio
of the
$\lb \rightarrow \psi(2S) \Lambda^0$ and
$\lb \rightarrow J/\psi \Lambda^0$ decays has been measured to be
$\Gamma(\lb \rightarrow \psi(2S)\Lambda^0)/\Gamma(\lb \rightarrow J/\psi\Lambda^0) = 0.501\pm 0.033 ({\rm stat})\pm 0.016({\rm syst})\pm 0.011 ({\bran})$.
The ratio
falls into the range 0.5--0.8, as found for the branching ratios
of analogous $B$ meson decays~\cite{pdg2014}. 
The only available theoretical expectation for the branching ratio of the two $\lb$ decays
($0.8\pm 0.1$~\cite{gutsche}) exceeds the measured value.

\section*{Acknowledgements}
% Acknowledgements for papers with collision data
% Version 23-Mar-2015

% Standard acknowledgements start here
%----------------------------------------------
We thank CERN for the very successful operation of the LHC, as well as the
support staff from our institutions without whom ATLAS could not be
operated efficiently.

We acknowledge the support of ANPCyT, Argentina; YerPhI, Armenia; ARC,
Australia; BMWFW and FWF, Austria; ANAS, Azerbaijan; SSTC, Belarus; CNPq and FAPESP,
Brazil; NSERC, NRC and CFI, Canada; CERN; CONICYT, Chile; CAS, MOST and NSFC,
China; COLCIENCIAS, Colombia; MSMT CR, MPO CR and VSC CR, Czech Republic;
DNRF, DNSRC and Lundbeck Foundation, Denmark; EPLANET, ERC and NSRF, European Union;
IN2P3-CNRS, CEA-DSM/IRFU, France; GNSF, Georgia; BMBF, DFG, HGF, MPG and AvH
Foundation, Germany; GSRT and NSRF, Greece; RGC, Hong Kong SAR, China; ISF, MINERVA, GIF, I-CORE and Benoziyo Center, Israel; INFN, Italy; MEXT and JSPS, Japan; CNRST, Morocco; FOM and NWO, Netherlands; BRF and RCN, Norway; MNiSW and NCN, Poland; GRICES and FCT, Portugal; MNE/IFA, Romania; MES of Russia and NRC KI, Russian Federation; JINR; MSTD,
Serbia; MSSR, Slovakia; ARRS and MIZ\v{S}, Slovenia; DST/NRF, South Africa;
MINECO, Spain; SRC and Wallenberg Foundation, Sweden; SER, SNSF and Cantons of
Bern and Geneva, Switzerland; NSC, Taiwan; TAEK, Turkey; STFC, the Royal
Society and Leverhulme Trust, United Kingdom; DOE and NSF, United States of
America.

The crucial computing support from all WLCG partners is acknowledged
gratefully, in particular from CERN and the ATLAS Tier-1 facilities at
TRIUMF (Canada), NDGF (Denmark, Norway, Sweden), CC-IN2P3 (France),
KIT/GridKA (Germany), INFN-CNAF (Italy), NL-T1 (Netherlands), PIC (Spain),
ASGC (Taiwan), RAL (UK) and BNL (USA) and in the Tier-2 facilities
worldwide.
%----------------------------------------------

%\section{Results}
%Place your content here
%
%\section{Conclusion}
%Place your content here
%%%%%%%%%%%%%%%%%%%%%%%%%%%%%%%%%%%%%%%%%%%%%%%%%%%%%%%%%%%%%%%%%%%%%%%%%%%%%%%
% Bibliography
%%%%%%%%%%%%%%%%%%%%%%%%%%%%%%%%%%%%%%%%%%%%%%%%%%%%%%%%%%%%%%%%%%%%%%%%%%%%%%
%
% % Style file to use with mcite.
% % Use atlasstyle with just cite.
% \newpage
% \bibliographystyle{atlasstylem}
% \bibliography{TrkEffNote}

\newpage
% ATLAS Collaboration author list
% Data extracted on 11-Aug-2015 for paper reference BPHY-2013-08
%\documentclass[11pt]{article}
%\usepackage{a4wide}\begin{document}
\begin{flushleft}
{\Large The ATLAS Collaboration}

\bigskip

G.~Aad$^{\rm 85}$,
B.~Abbott$^{\rm 113}$,
J.~Abdallah$^{\rm 151}$,
O.~Abdinov$^{\rm 11}$,
R.~Aben$^{\rm 107}$,
M.~Abolins$^{\rm 90}$,
O.S.~AbouZeid$^{\rm 158}$,
H.~Abramowicz$^{\rm 153}$,
H.~Abreu$^{\rm 152}$,
R.~Abreu$^{\rm 116}$
% ...
\bigskip
\\
% ...
$^{11}$ Institute of Physics, Azerbaijan Academy of Sciences, Baku, Azerbaijan\\
% ...
$^{85}$ CPPM, Aix-Marseille Universit{\'e} and CNRS/IN2P3, Marseille, France\\
% ...
$^{90}$ Department of Physics and Astronomy, Michigan State University, East Lansing MI, United States of America\\
% ...
$^{107}$ Nikhef National Institute for Subatomic Physics and University of Amsterdam, Amsterdam, Netherlands\\
% ...
$^{113}$ Homer L. Dodge Department of Physics and Astronomy, University of Oklahoma, Norman OK, United States of America\\
% ...
$^{116}$ Center for High Energy Physics, University of Oregon, Eugene OR, United States of America\\
% ...
$^{151}$ Institute of Physics, Academia Sinica, Taipei, Taiwan\\
$^{152}$ Department of Physics, Technion: Israel Institute of Technology, Haifa, Israel\\
$^{153}$ Raymond and Beverly Sackler School of Physics and Astronomy, Tel Aviv University, Tel Aviv, Israel\\
% ...
$^{158}$ Department of Physics, University of Toronto, Toronto ON, Canada\\
% ...
\end{flushleft}

%\end{document}
% Created with ./xml2latex.py

\end{document}